\documentclass[%
 pra,reprint,
 amsmath,amssymb,
 aps,
floatfix,
]{revtex4-2}

\usepackage{graphicx}
\usepackage{dcolumn}
\usepackage{bm}
\usepackage{siunitx}
\usepackage{physics}
\usepackage{xcolor}
\usepackage[caption=false]{subfig}
\usepackage[export]{adjustbox}

\usepackage{hyperref}
\hypersetup{
        colorlinks=false
}

\usepackage{nicematrix}
\NiceMatrixOptions{
code-for-first-row = \color{black} ,
code-for-last-row = \color{black} ,
code-for-first-col = \color{black} ,
code-for-last-col = \color{black}
}

\AtBeginDocument{\RenewCommandCopy\qty\SI} 
\begin{document}

\preprint{APS/123-QED}

\title{Resonant Schrödinger Cat States in Circuit Quantum Electrodynamics}

\author{M. Ayyash}\thanks{mmayyash@uwaterloo.ca}

\author{X. Xu}

\author{M. Mariantoni}
\thanks{matteo.mariantoni@uwaterloo.ca}

\affiliation{%
 Institute for Quantum Computing, University of Waterloo, 200 University Avenue West, Waterloo,
Ontario N2L 3G1, Canada\\
Department of Physics and Astronomy, University of Waterloo, 200 University Avenue West, Waterloo,
Ontario N2L 3G1, Canada
}%

\date{\today}

\begin{abstract}
We propose a fast scheme to generate Schrödinger cat states in a superconducting resonator using a continuously driven qubit without resorting to the dispersive regime, two-photon drives, or engineered two-photon dissipation. We provide analysis for when the qubit is on and off resonance from the drive. We extend our analysis to account for a third level in a weakly-anharmonic qutrit. We also discuss the case of a strongly-anharmonic qutrit. Throughout the paper, we corroborate our analytical results with numerical simulations in the presence of energy relaxation and dephasing of the qubit and resonator using realistic experimental parameters.

\end{abstract}

\maketitle


\section{\label{sec:level1} Introduction}
Quantum theory is driving an innovative transformation that extends beyond the realms of physics, profoundly impacting diverse fields such as information processing, metrology, and communication. Notably, quantum information processing and computing exhibit capabilities that far exceed those of classical devices. This advantage stems from the unique, non-classical characteristics inherent in quantum systems, which cannot be emulated by devices operating solely on classical principles.

Non-classical states, such as cat states, have played a significant role in quantum theory due to their incorporation of two main properties: superposition and entanglement. These states are valuable resources for information processing and computing. Cat states, specifically, are a quantum superposition of two well-defined macroscopic states and are inspired by Erwin Schrödinger's famous thought experiment involving a classical macroscopic system entangled with a quantum microscopic system \cite{CatParadox}.

{In the field of quantum optics, cat states refer to a phenomenon in which an electromagnetic field mode exists in a superposition of two diametrically opposed states. Typically, the superposed states are well-separated coherent states. Here, we specify cat states as the superposition of opposite-phase coherent states.} These states have experienced a recent resurgence in interest for two main reasons. Firstly, cat states provide a valuable resource for investigating fundamental aspects of quantum mechanics and pushing the boundaries of quantum coherence. As the size of the cat state grows, moving from mesoscopic or microscopic to macroscopic physics, it allows for the exploration of foundational questions. This growth in size, measured by the relative separation of the two coherent states, enables researchers to study the limits of quantum coherence \cite{ZurekDecoherence}. Secondly, cat states have generated interest in the development of error correction codes for quantum computation. They can be used to encode logical qubits, making amplitude damping or photon loss errors more easily correctable \cite{CochraneCat}.

Circuit quantum electrodynamics (QED) has emerged as a valuable testbed for investigating intriguing light-matter interaction phenomena that are othwerwise inaccessible to researchers relying on natural atoms and light sources \cite{cQEDRev}. By exploring the interaction between qubits and quantized microwave photons, circuit QED offers a versatile platform for generating cat states within superconducting resonators.

{The generation of cat states has been extensively studied theoretically \cite{SA_Nonclassical,QR_modulation1,qOptical_9,SA_Rephasing,KerrTL,Referee_cQED,QR_modulation2,CatIonTheory,NonlinearOpticsCat} and has seen numerous experimental demonstrations in various platforms \cite{qOpticalCat_1,qOpticalCat_2,qOpticalCat_3,qOpticalCat_4,qOpticalCat_5,qOptical_6,qOptical_7,qOptical_8,CatIon,CatIon2,QR_Haroche}. These experimental demonstrations encompass the realization of cat states in microwave fields \cite{HarochExp}, the manipulation of motional degrees of freedom in trapped ions \cite{CatIon}, and the creation of phononic cats in a mechanical resonator \cite{OptoMechCat}.}

In the context of circuit QED, the most pertinent methods for generating cat states are listed here. A Kerr-nonlinear resonator can be initialized in a coherent state, then time-evolved and measured at specific times 
 to create a cat state \cite{YurkeAmpDispCat}. A cat state can be synthesized one Fock state at the time by using an excited ancillary qubit, which loads the resonator by performing resonant Rabi
swaps \cite{HofheinzSynth}. An ancillary qubit can also be used in the strong dispersive regime along with resonator-dependent qubit pulses and qubit-dependent resonator displacements \cite{qcMAP,qcMAPexp}. A dissipation-based approach can yield cat states as the steady state of a system driven by a two-photon process along with an engineered two-photon loss \cite{QO_Driv2Ph,QO_Disp2Ph,TwoPhotonLossCat}. Lastly, cat states can be generated and stabilized using a Kerr-nonlinear resonator driven with a parametric two-photon drive \cite{TwoPhotonDrivCat,KerrCatExp}. However, in order to fully realize the potential of a bosonic cat code \cite{BosonicQIPRev}, it is necessary to establish a reliable technique for generating cat states on time scales shorter than those required in all the aforementioned methods.

In this paper, we propose an approach for producing cat states in a superconducting resonator building upon a previous cavity QED proposal \cite{SolanoCat}. We significantly extend the analysis offered in that proposal, clarify a few pitfalls inherent to that proposal, and examine examples with realistic parameters based on a circuit QED implementation. Our method involves a driven qubit-resonator system, with the driving focused exclusively on the ancillary qubit. This driving can be either resonant or detuned, although we find that resonant driving yields larger cat states. By implementing our method, the photon number, which coincides with the size of the cat state, grows quadratically over time when the drive and resonator are perfectly resonant. This quadratic growth allows for a major speed-up compared to alternative methods using identical setup parameters to generate a cat state. Furthermore, our method reliably produces cat states with the parity depending on the qubit state. It is important to note that the parity (even or odd) of the cat state is contingent upon the state of the qubit. Overall, our approach offers a promising avenue for generating cat states in superconducting resonators efficiently and reliably.

The paper outline is as follows. In Sec.~\ref{sec:2}, we present the theoretical framework for a resonant and detuned driven qubit-resonator system. We derive the precise conditions necessary to achieve a cat state in the resonator. In Sec.~\ref{sec:4}, we extend the theory to account for a three-level system (qutrit) of varying anharmonicity. We discover a dark state in the weakly-anharmonic regime, resulting in another non-classical state in the resonator: a superposition of a vacuum state and a cat state. We also discuss the use of the qubit vs. qutrit analysis depending on the anharmonicity of the implementation scheme, i.e., whether a circuit is closer to a qubit or qutrit (e.g., charge qubit vs. transmon qubit). Additional experimental considerations, such as decoherence and spurious driving of the resonator, are discussed in App.~\ref{app:Decoherence} and~\ref{sec:Spurious}, respectively. In Sec.~\ref{sec:5}, we discuss our method in comparison with other widespread methods. We provide a summary and concluding remarks in Sec.~\ref{sec:6}. 
 
\section{Driven Qubit-Resonator System}\label{sec:2}

The interaction between a qubit and a resonator can be effectively described by the quantum Rabi model (QRM), which operates under the electric dipole and single mode approximations. In our study, we specifically investigate the QRM in the presence of a continuous qubit drive. It is important to note that we choose to operate our system in a parameter regime where the usual rotating wave approximation (RWA), which is commonly used to obtain the Jaynes-Cummings model (JCM), cannot be employed. This approach allows us to explore a different aspect of the qubit-resonator interaction and gain further insights into its behavior. By adding a drive to the qubit, we arrive at the driven QRM. The drive dresses the qubit frequency and, as a result, it affects its interaction with the resonator. The drive parameters now play a crucial role in determining the regimes where an RWA is applicable.

In Sec.~\ref{sec:QubitSysHamCond}, we begin by stating the system Hamiltonian and establishing the necessary conditions for implementing an RWA within the framework of qubit-drive resonant conditions. In Sec.~\ref{sec:RWAHam}, we delve into the RWA Hamiltonian and explore how a second, distinct RWA can be imposed under strong driving conditions. Furthermore, we explore the application of this approach in generating a cat state and encoding a qubit state in the resonator. We discuss the effect of the neglected counter-rotating driving and interaction terms manifesting in the Bloch-Siegert shifts. In Sec.~\ref{sec:DetunedTheory}, we expand the theory to cover arbitrary detuning scenarios. We discover that it remains feasible to generate a cat state, similar to the resonant case, albeit with a reduced number of photons. Throughout the section, we corroborate the robustness of our predictions by numerically solving a master equation that accounts for qubit relaxation, dephasing, and resonator relaxation described in App.~\ref{app:Decoherence}.


\subsection{RWA Hamiltonian}\label{sec:QubitSysHamCond}
We start by considering the Hamiltonian associated with the driven QRM that reads as
 \begin{align}
     \widehat{H}= &\frac{\hbar \omega_{\text{q}}}{2}\hat{\sigma}_z + \hbar \omega_{\text{r}} \hat{a}^\dagger \hat{a} + \hbar g (\hat{\sigma}_+ + \hat{\sigma}_-)(\hat{a}^\dagger + \hat{a}) \nonumber\\ &+ \hbar\Omega\cos(\omega_{\text{d}} t)(\hat{\sigma}_+ + \hat{\sigma}_-),
     \label{eq:SysHam}
 \end{align}
where $\hat{\sigma}_z=\dyad{\text{e}}-\dyad{\text{g}}$ describes the population difference between the excited energy state $\ket{\text{e}}$ and the ground state $\ket{\text{g}}$ of the qubit, $ \hat{\sigma   }_+=\dyad{\text{e}}{\text{g}}$ and $\hat{\sigma}_-=\hat{\sigma}_+^\dagger$ are the raising and lowering operators of the qubit, $\hat{a}$ and $\hat{a}^\dagger$ are the annihilation and creation operators of the resonator,  $\omega_{\text{q}}$ is the transition (angular) frequency of the qubit, $\omega_{\text{r}}$ is the resonance (angular) frequency of the resonator, $g$ is the coupling strength between the qubit and resonator, $\Omega$ is the strength of the classical field, $\omega_{\text{d}}$ is the classical field driving frequency, and $t$ is time.

The Hamiltonian of Eq.\eqref{eq:SysHam} can be written in the frame of the driving field by means of the unitary transformation $\hat{U}=\exp[-i\omega_{\text{d}}t(\hat{\sigma}_z/2 + \hat{a}^\dagger \hat{a})]$, 
\begin{align}
    \widehat{H}^{\text{d}} &=\hat{U}^\dagger \widehat{H} \hat{U} +i\hbar \dot{\hat{U}}^\dagger\hat{U}\nonumber \\ &=\frac{\hbar \Delta}{2}\hat{\sigma}_z + \hbar \delta\hat{a}^\dagger\hat{a}\nonumber \\ &\,\,\,\,\,\,\,+ \hbar g\left( \hat{\sigma}_+\hat{a} +\hat{\sigma}_-\hat{a}^\dagger +e^{+i2\omega_{\text{d}}t}\hat{\sigma}_+\hat{a} ^\dagger +e^{-i2\omega_{\text{d}}t}\hat{\sigma}_-\hat{a}\right)\nonumber \\ &\,\,\,\,\,\,\,+ \frac{\hbar \Omega}{2}\left(\hat{\sigma}_+ + \hat{\sigma}_- +e^{+i2\omega_{\text{d}}t}\hat{\sigma}_+ + e^{-2i\omega_{\text{d}}t}\hat{\sigma}_-\right),
    \label{eq:SysHamRotFr}
\end{align}
where $\Delta=\omega_{\text{q}} -\omega_{\text{d}}$, $\delta=\omega_{\text{r}} - \omega_{\text{d}}$, and we use the exponential definition of $\cos(x)=(e^{ix}+e^{-ix})/2$.

The Hamiltonian of Eq.~\eqref{eq:SysHamRotFr} can be simplified by imposing a set of RWA conditions that read as
\begin{subequations}\label{eq:RWAConditions}
\begin{align}
    &\omega_{\text{q}}-\omega_{\text{r}} \ll \omega_{\text{q}} +\omega_{\text{r}} \text{ and } g \ll \min(\omega_{\text{q}},\omega_{\text{r}}),\label{eq:RWAConditions1} \\
    &g\ll 2\omega_{\text{d}}, \text{ and}\label{eq:RWAConditions2}\\
    &\Omega\ll 4\omega_{\text{d}}.
    \label{eq:RWAConditions3}
\end{align}
\end{subequations}
The conditions in Eq.~\eqref{eq:RWAConditions1} are those used to derive the JCM from the QRM. The other two conditions allow us to account for the presence of the driving field. The condition in Eq.~\eqref{eq:RWAConditions2} is necessary to eliminate the counter-rotating interaction terms $g(e^{+i2\omega_{\text{d}}t}\hat{\sigma}_+\hat{a} ^\dagger +e^{-i2\omega_{\text{d}}t}\hat{\sigma}_-\hat{a})$, whereas the condition in Eq.~\eqref{eq:RWAConditions3} is required to drop the counter-rotating driving terms $\Omega(e^{+i2\omega_{\text{d}}t}\hat{\sigma}_+ +e^{-i2\omega_{\text{d}}t}\hat{\sigma}_-)/2$. Under all these RWA conditions, the Hamiltonian can be simplified to read as
\begin{align}
\widehat{H}^{\text{d}}_{\text{RWA}}=\,&\frac{\hbar \Delta}{2}\hat{\sigma}_z + \hbar \delta\hat{a}^\dagger\hat{a} + \hbar g( \hat{\sigma}_+\hat{a} +\hat{\sigma}_-\hat{a}^\dagger) \nonumber\\
    &+ \frac{\hbar \Omega}{2}( \hat{\sigma}_+ +\hat{\sigma}_-).
    \label{eq:RWAHam}
\end{align}
This Hamiltonian, which is free of any time-dependent terms, serves as the basis for the work presented in this section.

\subsection{Resonant strong driving regime}\label{sec:RWAHam}

\begin{figure*}[t]
        \includegraphics[scale=1,left]{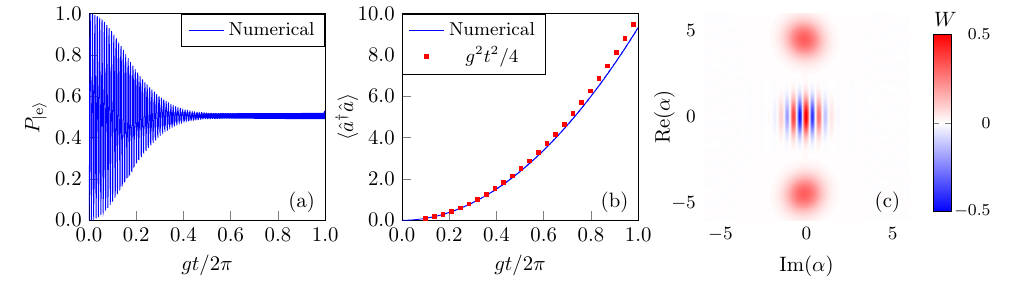}
        \caption{Characterization of cat states under resonance conditions. All the numerical simulations of the master equation are performed with QuTiP \cite{Johansson_Nation_Nori_2012,Johansson_Nation_Nori_2013}. The parameters used for the simulations are: $\Omega=2\pi\times \SI{2}{\giga\hertz}$. $\Delta=\delta=0$. $\omega_{\text{q}}=2\pi\times \SI{5}{\giga\hertz}$. $g=2\pi \times \SI{20}{\mega\hertz}$. $\gamma_1=\kappa=\SI{500}{\kilo\hertz}$. $\gamma_\phi= \SI{1}{\mega\hertz}$.   (a),(b) $P_{\ket{\text{e}}}$ and $\langle \hat{a}^\dagger \hat{a}\rangle$ vs. normalized time $gt/2\pi$. (c) Heatmap of the Wigner function $W$ at $gt/2\pi= 1.0$ when the qubit is measured in $\ket{\text{e}}$. This choice of time corresponds to an ideal cat state, accounting for the driving Bloch-Siegert shift. The simulations are performed in the laboratory frame and, thus, the resonator state rotates with $e^{-i\omega_{\text{r}}t}$.}
        \label{fig:QubitResDynamics}
\end{figure*}
\twocolumngrid

We start by considering the Hamiltonian of Eq.\eqref{eq:RWAHam} under the resonant condition $\Delta=0$. To simplify the notation, we define $\widehat{H}_0^{\text{d}}=\hbar \Omega (\hat{\sigma}_+ + \hat{\sigma}_-)/2 + \hbar \delta\hat{a}^\dagger\hat{a}$ and $\widehat{H}_{\text{I}}^{\text{d}}=\hbar g\left( \hat{\sigma}_+\hat{a} +\hat{\sigma}_-\hat{a}^\dagger \right)$. We then apply the unitary transformation $\hat{U}_0=\exp(-i \widehat{H}_0 t /\hbar)$, which allows us to obtain the interaction picture Hamiltonian
\begin{align}
    \widehat{H}_{\text{RWA}}^{\text{(I)}} = \hat{U}_0^\dagger \widehat{H}_{\text{I}}\hat{U}_0 = & \frac{\hbar g}{2} \bigg(\dyad{+}-\dyad{-}\nonumber \\ &+ e^{i \Omega t} \dyad{+}{-} - e^{-i\Omega t}\dyad{-}{+}\bigg)\hat{a}e^{-i\delta t} \nonumber \\ &+ \text{H.c.};
    \label{eq: IntPicRWA}
\end{align}
$\ket{\pm}$ are the dressed basis qubit eigenstates, with the property that $\hat{\sigma}_x \ket{\pm}=\pm\ket{\pm}$.

The Hamiltonian of Eq.~\eqref{eq: IntPicRWA} reveals two distinct interactions taking place at different timescales. In one of them, two terms are modulated by the driving field and exhibit oscillations with functional dependence $e^{\pm i \Omega t}$. These terms can be neglected by imposing the \textit{strong driving condition}, 
\begin{align}
    \Omega \gg |\delta|,g.
\end{align}
This condition, however, must be considered at the same time as the condition of Eq.\eqref{eq:RWAConditions3}; therefore, the complete condition reads as
\begin{align}\label{eq:FullStrongDrivCond}
    g,|\delta| \ll \Omega \ll 4 \omega_{\text{d}}.
\end{align}
This means that $\Omega$ is characterized by a lower bound as well as an upper bound, that is, it cannot be arbitrarily large \footnote{The condition in  Eq.~\eqref{eq:FullStrongDrivCond} is the correction needed to make the theory of Ref.~\cite{SolanoCat} work for a driven qubit-resonator in circuit QED.}. This assumption makes it possible to obtain the effective Hamiltonian
\begin{align}\label{eq:CatHam}
    \widehat{{H}}_{\text{eff}}^{\text{(I)}}= \frac{\hbar g}{2}\left( \dyad{+}-\dyad{-}\right)\left(\hat{a}^\dagger e^{+i\delta t}+\hat{a}e^{-i\delta t}\right).
\end{align}
The dynamics associated with this Hamiltonian result in a conditional displacement of the resonator state based on the qubit state. Specifically, if the qubit is in state $\ket{+}$, the resonator state is displaced in a certain direction. Conversely, if the qubit state is in $\ket{-}$, the resonator state is displaced in the opposite direction. {Explicitly, the time-evolution operator generated by this Hamiltonian is \footnote{Note that, unlike our method, the displacements using typical dispersive methods are conditioned on the bare basis $\{\ket{\text{g}},\ket{\text{e}}\}$.}
\begin{align}
 \widehat{U}_{\text{eff}}(t,0)=\dyad{+}\widehat{D}(\alpha) + \dyad{-}\widehat{D}(-\alpha),   
\end{align}
where $\widehat{D}(\alpha)=\exp(\alpha \hat{a}^\dagger - \alpha^* \hat{a})$ is the displacement operator and $\alpha=-g(e^{i\delta t}-1)/2\delta$; when $\delta\rightarrow 0$, then $\alpha=-igt/2$.}
Thus, if we choose the initial state to be $\ket{\psi_{\text{i}}}=\ket{\text{g}}\ket{0}=(\ket{+}+\ket{-})\ket{0}/\sqrt{2}$, the time evolution of the system leads, in the interaction picture, to the state
\begin{align}
    \ket{\psi(t)}^{\text{(I)}}&=\frac{1}{\sqrt{2}}(\ket{+}\ket{\alpha}+\ket{-}\ket{-\alpha}) \nonumber\\
    &=\frac{1}{2}\ket{\text{g}}(\ket{\alpha}+\ket{-\alpha}) + \frac{1}{2}\ket{\text{e}}(\ket{\alpha}-\ket{-\alpha}).
    \label{eq:CatState}
\end{align}

If the qubit is measured to be in $\ket{\text{g}}$, the resonator is left in a state that is proportional to the superposition of two coherent states, $\bra{\text{g}}\ket{\psi(t)}^{\text{(I)}}\propto\ket{\alpha} + \ket{-\alpha}$; this state is commonly referred to as an \emph{even cat state}. On the other hand, if the qubit is measured in $\ket{\text{e}}$, the resonator is in the state $\bra{\text{e}}\ket{\psi(t)}^{\text{(I)}}\propto\ket{\alpha} - \ket{-\alpha}$; this state is known as an \emph{odd cat state}.

Using this procedure, we can encode an arbitrary state of a qubit, $\ket{\psi_{\text{q}}}=c_{\text{g}}\ket{\text{g}}+c_{\text{e}}\ket{\text{e}}$ (with $|c_{\text{g}}|^2+|c_{\text{e}}|^2=1$), into the resonator via cat states. Let the initial state be $\ket{\psi_{\text{i}}}=(c_{\text{g}}\ket{+} + c_{\text{e}}\ket{-})\ket{0}$. After time-evolving for the desired period and measuring in the bare basis $\{\ket{\text{g}},\ket{\text{e}}\}$, the resonator is left in a state $\propto c_{\text{g}}\ket{\alpha}\pm c_{\text{e}}\ket{-\alpha}$. This encoding is an instance of a bosonic logical qubit encoding using two-component cat states \cite{qcMAP,BosonicQIPRev}.

Figure~\ref{fig:QubitResDynamics} displays the results of numerical simulations of the complete system Hamiltonian of Eq.~(\ref{eq:SysHam}), without any approximations. These simulations are performed in presence of both qubit and resonator decoherence by means of a Lindblad master equation, as explained in App.~\ref{app:Decoherence}. Figure~\ref{fig:QubitResDynamics}~(a) shows the probability of the qubit to be in the excited state, $P_{\ket{\text{e}}}$. This probability shows that as the cat state grows, the qubit population converges to an equal superposition of $\ket{ \text{g}}$ and $\ket{\text{e}}$. Figure~\ref{fig:QubitResDynamics}~(a) indicates that the photon number $n=\langle \hat{a}^{\dag} \hat{a} \rangle$ grows quadratically in time. Interestingly, this behaviour persists even in presence of decoherence. The numerical results closely follow the analytical prediction $|\alpha|^2=g^2 t^2 /4$.

In order to visualize the cat states, we elect to represent them in phase space by means of the Wigner function \cite{EMMPhaseSpaceQM}
\begin{align}
    W(\alpha,\alpha^*)=\frac{1}{\pi\hbar}\Tr\left(\widehat{D}(2\alpha) e^{i\pi \hat{a}^\dagger \hat{a}}\hat{\rho}_{\text{r}}\right),
\end{align}
where $\hat{\rho}_\text{r}$ is the resonator density matrix \footnote{in our case, $\hat{\rho}_{\text{r}}$ is obtained after performing a qubit state projective measurement}. Figure~\ref{fig:QubitResDynamics}(c) shows $W$ after measuring the qubit for $gt/2\pi=1.0$.

The numerical simulations based on the Hamiltonian of Eq.~\eqref{eq:SysHam}  do not employ the approximations of Eqs.~\eqref{eq:RWAConditions} and \eqref{eq:FullStrongDrivCond}. This exact Hamiltonian can be rewritten in the rotating frame and interaction picture as
\begin{align}
        \widehat{H}^{(\text{I})}=& \frac{\hbar }{2}\bigg(\dyad{+}-\dyad{-}\nonumber\\ & + e^{i \Omega t} \dyad{+}{-} - e^{-i\Omega t}\dyad{-}{+}\bigg)\nonumber\\ &\times \left(   \underbrace{\frac{\Omega}{2}e^{i2\omega_{\text{d}}t} }_{\substack{\text{driving}\\ \text{Bloch-Siegert shift}}} 
  + \overbrace{g\hat{a}^\dagger e^{i(2\omega_{\text{d}} + \delta )t}}^{\substack{\text{interaction}\\ \text{Bloch-Siegert shift}}} + g\hat{a} e^{-i\delta t}\right)\nonumber \\&+ \text{H.c.},
 \label{eq:FullIntPic}
\end{align}
which includes the counter-rotating driving and interaction terms. These terms lead to the driving and interaction Bloch-Siegert shifts, respectively. These two effects combined together result in the oscillation of the amplitude and weight coefficients of the cat states' lobes \cite{StrongDrivBlochSiegShift}. We are only working in the strong coupling regime where the interaction Bloch-Siegert shift is negligible. In the Supplementary Materials of Ref.~[tbd], we show a movie comparing side-by-side the approximated analytical and exact numerical simulations of the cat state evolution. Except for the oscillations due to the driving Bloch-Siegert shift, the two solutions very closely resemble each other. It is worth noting that these oscillations can be tracked deterministically, therefore allowing us to measure an ideal cat state.

\subsection{Detuned regime}\label{sec:DetunedTheory}
Thus far, our analysis has been confined to the resonance condition where $\Delta=0$. However, to achieve scalability, it is essential to explore the generation of cat states without relying solely on qubit-drive resonance. One possible application of our theory is the generation of cat states in multiple driven qubit-resonator systems. In pursuit of this goal, we extend the theory to the qubit-drive detuned regime, where $\Delta \neq 0$.

We consider the Hamiltonian of Eq.~\eqref{eq:FullIntPic} when $\Delta\neq 0$. In this case,
\begin{align}
    \widehat{H}^{(\text{I})}_{\text{det}}=& \frac{\hbar }{2\varepsilon(\Delta+\varepsilon)}\nonumber\\ &\times\bigg[ (\Delta+\varepsilon)\Omega \dyad{\widetilde{+}}+(\Delta+\varepsilon)^2e^{i\varepsilon t}\dyad{\widetilde{+}}{\widetilde{-}}\nonumber\\ &- \Omega^2 e^{-i \varepsilon t}\dyad{\widetilde{-}}{\widetilde{+}} - (\Delta+\varepsilon)\Omega\dyad{\widetilde{-}}\bigg]\nonumber\\ &\times \left(   \frac{\Omega}{2}e^{i2\omega_{\text{d}}t}   + g\hat{a} e^{-i\delta t} + g\hat{a}^\dagger e^{i(2\omega_{\text{d}} + \delta )t}\right) + \text{H.c.}; 
\end{align} $\varepsilon=\sqrt{\Omega^2 + \Delta^2}$ and the detuned qubit orthonormal basis is defined as $ \ket{\widetilde{+}}=\sin\left(\theta /2\right)\ket{\text{g}}+\cos\left(\theta /2\right)\ket{\text{e}} $ and $\ket{\widetilde{-}}=\cos\left(\theta /2\right)\ket{\text{g}}-\sin\left(\theta /2\right)\ket{\text{e}} $,
where the mixing angle $\theta= \arctan(\Omega/\Delta).$ Note that, if we set $\Delta=0$, we recover the results of Sec.~\ref{sec:RWAHam}. If $\Omega \gg \Delta$, then $\ket{\widetilde{+}}\simeq \ket{+}$ and $\ket{\widetilde{-}}\simeq \ket{-}$; on the other hand, if $\Delta \gg \Omega$, we have $\ket{\widetilde{+}}\simeq \ket{\text{e}}$ and $\ket{\widetilde{-}}\simeq \ket{\text{g}}$ .

 We now follow a similar procedure as in Sec.~\ref{sec:RWAHam}  but in presence of detuning. Assuming all the conditions of Eq.~\eqref{eq:RWAConditions} to hold true and additionally imposing the \textit{strong driving-detuning} condition
\begin{align}
g,|\delta|\ll\varepsilon\ll 
 4 \omega_{\text{d}},    
\end{align}
we perform an RWA obtaining the effective Hamiltonian
\begin{align}
        \widehat{H}^{\text{ (I)}}_{\text{det, eff}}  = \frac{\hbar g \Omega}{2\varepsilon}\left(\dyad{\widetilde{+}} - \dyad{\widetilde{-}} \right)\left(\hat{a}^\dagger e^{+i\delta t}+\hat{a}e^{-i\delta t}\right).
\end{align}
\begin{figure*}[t]
        \includegraphics[scale=1,left]{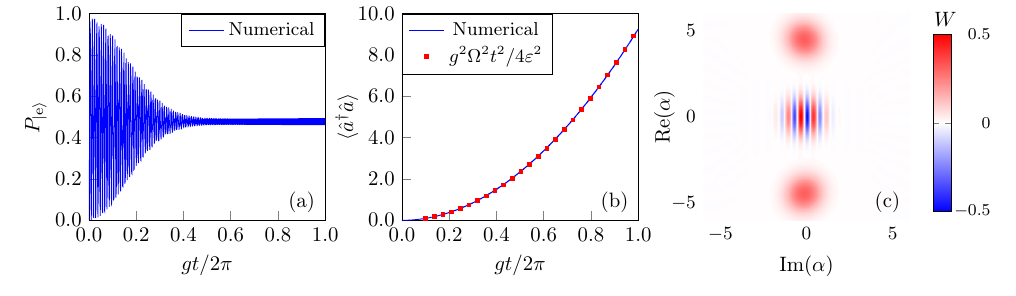}
        \caption{Characterization of cat states under detuned conditions. The parameters used for these simulations are identical to those used in Fig.~\ref{fig:QubitResDynamics}, except for $\Delta=2\pi \times \SI{500}{\mega\hertz}$.   (a),(b) $P_{\ket{\text{e}}}$ and $\langle \hat{a}^\dagger \hat{a}\rangle$ vs. normalized time $g t/2\pi$. (c) Heatmap of the Wigner function $W$ at $gt/2\pi= 1.0$. By selecting this specific time, an ideal cat state is achieved, characterized by equal weight coefficients. This outcome is attained by leveraging the oscillations resulting from the driving Bloch-Siegert shift.}
        \label{fig:QubitDetDynamics}
\end{figure*}
\twocolumngrid
This Hamiltonian generalizes that of Eq.~\eqref{eq:CatHam} and, thus, it generates a resonator displacement conditioned on $\ket{\widetilde{\pm}}$. When initializing the system in state $\ket{\psi_{\text{i}}}=\ket{\text{g}}\ket{0}=(\sin\left(\theta/2\right)\ket{\widetilde{+}}+\cos\left(\theta/2\right)\ket{\widetilde{-}})\ket{0}$, we obtain the time-evolved state 
\begin{align}\label{eq:DetunedCatInt}
    \ket{\psi(t)}^{\text{(I)}}&=\left(\sin\left(\frac{\theta}{2}\right)\ket{\widetilde{+}}\ket{\widetilde{\alpha}}+\sin\left(\frac{\theta}{2}\right)\ket{\widetilde{-}}\ket{-\widetilde{\alpha}} \right)\nonumber \\
    &= \Bigg[\ket{\text{g}}\left( \frac{1}{2}\sin\left(\theta\right)\ket{\widetilde{\alpha}}+ \sin^2\left(\frac{\theta}{2}\right)\ket{-\widetilde{\alpha}}\right)\nonumber \\
    &\,\,\,\,\,\,\,+ \ket{\text{e}}\left( \frac{1}{2}\sin\left(\theta\right) \ket{\widetilde{\alpha}} - \cos^2\left(\frac{\theta}{2}\right)\ket{-\widetilde{\alpha}}\right)\Bigg],
\end{align}
where $\widetilde{\alpha}=-g\Omega(e^{i\delta t}-1)/2\varepsilon\delta$; when $\delta\rightarrow 0$, then $\widetilde{\alpha}=-ig\Omega t/2\varepsilon$. 
The state described by Eq.~\eqref{eq:DetunedCatInt} exhibits a superposition of coherent states with opposite phases, featuring distinct weight coefficients. This results in what we refer to as an \textit{asymmetrically weighted cat state}. Remarkably, cat states can be generated even when $\Delta\neq 0$, without necessarily requiring the system to be in the strong dispersive regime or resonant qubit-drive regime. However, it is essential to note that cat states produced in the detuned regime possess smaller amplitudes compared to those generated resonantly, even when subjected to time-evolution duration. This amplitude difference arises because $|\widetilde{\alpha}| <|\alpha|$. 

Under detuning conditions, the Bloch-Siegert and ac Stark shifts' oscillations can be harnessed to create an ideal cat state. This happens when the distinct weight coefficients become identical at specific times, resulting in the formation of the ideal cat state.

Figure~\ref{fig:QubitDetDynamics} displays the results of numerical simulations similar to those in Fig.~\ref{fig:QubitResDynamics} but under detuned conditions. The photon number behaves similarly to the resonant case, $|\widetilde{\alpha}|^2=g^2 \Omega ^2 t^2 / 4\varepsilon^2$, growing quadratically in time. This analytical result is validated by the numerical simulations.


\section{Extension to a Qutrit}\label{sec:4}
\begin{figure}[ht]
        \includegraphics[scale=1]{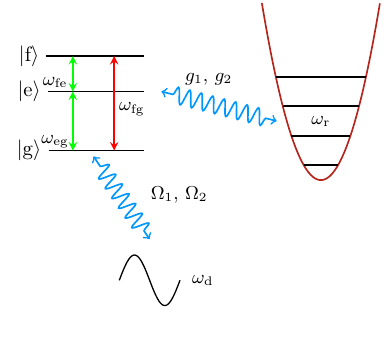}
        \caption{Schematic of a cascade ($\Xi$) type qutrit with both transitions simultaneously coupled to a resonator and a classical drive. A cascade qutrit has selection rules only permitting nearest neighbour transitions, i.e., $\ket{\text{g}}\leftrightarrow\ket{\text{e}}$ and $\ket{\text{e}}\leftrightarrow\ket{\text{f}}$ are allowed transitions, while $\ket{\text{g}}\leftrightarrow\ket{\text{f}}$ is forbidden. }\label{fig:QutritSchematic}
\end{figure}

In most implementation schemes, the device is rarely a true two-level system but rather infinite-dimensional. Depending on the degree of anharmonicity in a system, nearby transitions can significantly impact the system's dynamics. Many circuit QED implementations, including transmon, charge, and flux qubits (at the symmetry point), can be described as $\Xi$-type (cascade) qutrits when truncated to the first three levels (see Fig.~\ref{fig:QutritSchematic}). In this section, we extend our considerations to a driven $\Xi$-type qutrit interacting with a resonator, where both transitions are influenced by the drive and resonator.


In Sec.~\ref{sec:WeaklyAnharm}, we state the system Hamiltonian and derive the necessary RWA conditions for a weakly-anharmonic qutrit. We show that the cat-state-generating protocol can be successfully generalized with some modifications. We also generalize the mapping of a qubit state into the resonator state for the qutrit-resonator system. In Sec.~\ref{sec:StronglyAnharm}, we examine the qutrit in the strongly-anharmonic limit. While qubit considerations remain largely applicable, there are notable perturbations arising from leakage and interference associated with the third state. We discuss the case of an arbitrarily-anharmonic qutrit in App.~\ref{app:ArbAnharm}. 

\subsection{Extension to a weakly-anharmonic qutrit}\label{sec:WeaklyAnharm}

We start by generalizing the driven QRM to include a qutrit with two allowed transitions. The Hamiltonian for this system reads 
\begin{subequations}\label{eq:4-QutritSysHam}
 \begin{align}
    \widehat{H}= \widehat{H}_0 + \widehat{H}_{\text{I}} + \widehat{H}_{\text{d}},
\end{align} 
where
\begin{align}
    \widehat{H}_0 = \frac{\hbar \omega_{\text{eg}}}{2}\left(\dyad{\text{e}}-\dyad{\text{g}}\right)+ \frac{\hbar \widetilde{\omega}_{\text{f}}}{2}\dyad{\text{f}} + \hbar \omega_{\text{r}}\hat{a}^\dagger \hat{a},
\end{align}
\begin{align}
    \widehat{H}_{\text{d}}=&\hbar \cos(\omega_{\text{d}}t)\bigg[\Omega_1(\hat{\sigma}_{1+} + \hat{\sigma}_{1-} ) + \Omega_2(\hat{\sigma}_{2+} + \hat{\sigma}_{2-} ) \bigg],
\end{align}
and
\begin{align}
    \widehat{H}_{\text{I}}=&\hbar \bigg[g_1(\hat{\sigma}_{1+} + \hat{\sigma}_{1-} ) + g_2(\hat{\sigma}_{2+} + \hat{\sigma}_{2-} )\bigg] \left(\hat{a}^\dagger + \hat{a}\right).
\end{align}
\end{subequations}
Here, we define $\widetilde{\omega}_{\text{f}}=2\omega_{\text{fe}}+\omega_{\text{eg}}$, $\hat{\sigma}_{1+}:=\dyad{\text{e}}{\text{g}}$ ($\hat{\sigma}_{1-}=\hat{\sigma}_{1+}^\dagger$) as the raising (lowering) operator for the first transition, $\ket{\text{g}}\leftrightarrow\ket{\text{e}}$, with frequency $\omega_{\text{eg}}$ and $\hat{\sigma}_{2+}:=\dyad{\text{f}}{\text{e}}$ ($\hat{\sigma}_{2-}=\hat{\sigma}_{2+}^\dagger$) as the raising (lowering) operator for the second transition,  $\ket{\text{e}}\leftrightarrow\ket{\text{f}}$, with frequency $\omega_{\text{fe}}$. Also, $\Omega_1$ $(g_1)$ is the coupling strength between the drive (resonator) and the first transition, and $\Omega_2$ $(g_2)$ is the coupling strength between the drive (resonator) and the second transition.

We can rewrite the Hamitlonian of Eq.\eqref{eq:4-QutritSysHam} in a rotating frame by means of  the unitary transformation $\hat{U}=\exp[-it\left(\omega_{\text{d}}\left(\dyad{\text{e}}-\dyad{\text{g}}\right) + \widetilde{\omega}_{\text{f}}\dyad{\text{f}} + 2\omega_{\text{d}}\hat{a}^\dagger \hat{a}\right)/2] $,
\begin{align}\label{eq:4-QutritRotFrameFull}
    \widehat{H}^{\text{d}}=&\frac{\hbar \Delta_1}{2}\left(\dyad{\text{e}} - \dyad{\text{g}} \right)+ \hbar\delta \hat{a}^\dagger \hat{a}\nonumber\\
    &+  \frac{\hbar \Omega_1}{2}\left(\hat{\sigma}_{1+}+\hat{\sigma}_{1-}+e^{i2\omega_{\text{d}}t}\hat{\sigma}_{1+} +e^{-i2\omega_{\text{d}}t}\hat{\sigma}_{1-}\right)
    \nonumber\\ &+  \frac{\hbar \Omega_2}{2}\big(e^{i\widetilde{\chi} t/2}\hat{\sigma}_{2+} + e^{-i\widetilde{\chi} t/2}\hat{\sigma}_{2-}\nonumber \\ &+e^{i(\widetilde{\omega}_\text{f} +\omega_{\text{d}})t/2}\hat{\sigma}_{2+} + e^{-i(\widetilde{\omega}_\text{f} +\omega_{\text{d}})t/2}\hat{\sigma}_{2-}\big)\nonumber\\
    &+\hbar\bigg[ g_1\left(e^{i\omega_{\text{d}}t}\hat{\sigma}_{1+} + e^{-i\omega_{\text{d}}t}\hat{\sigma}_{1-}\right)\nonumber\\
    &+ g_2\left(e^{+i(\widetilde{\omega}_{\text{f}}-\omega_{\text{d}})t/2}\hat{\sigma}_{2+} + e^{-i(\widetilde{\omega}_{\text{f}}-\omega_{\text{d}})t/2}\hat{\sigma}_{2-}\right)\bigg]\nonumber \\&\times\left(e^{i\omega_{\text{d}}t}\hat{a}^\dagger + e^{-i\omega_{\text{d}}t}\hat{a} \right),
\end{align}
where $\Delta_1=\omega_{\text{eg}}-\omega_{\text{d}}$, $\delta=\omega_{\text{r}}-\omega_{\text{d}}$  and $\tilde{\chi}=\widetilde{\omega}_{\text{f}}-3\omega_{\text{d}}$. Henceforth, we set $\Delta_1=0$ so that the drive is on resonance with the first transition. Typically,  $\xi=\omega_{\text{fe}}-\omega_{\text{eg}}$ is the anharmonicity parameter between the first and second transitions. Depending on the energy level spacing of a given circuit, $\xi$ can be negative or positive. When $\Delta_1=0$, the parameter $\widetilde{\chi}$ becomes proportional to the anharmonicity $\xi$; $\widetilde{\chi}=\widetilde{\omega}_{\text{f}}-3 \omega_{\text{eg}}=2 \omega_{\text{fe}}-2\omega_{\text{eg}}=2\xi$.

For a qubit implementation like the transmon, the first and second transition frequencies are usually close and $\xi$ is negative. Additionally, the transition matrix elements of a transmon between the $n^{\text{th}}$ and $(n+1)^{\text{st}}$ states are proportional to $\sqrt{n+1}$. Thus, to simplify the analytical calculations, we assume a perfectly harmonic qutrit where $\widetilde{\chi}=0$, $g_2=\sqrt{2}g_1$ and $\Omega_2=\sqrt{2}\Omega_1$. We use this as a toy model to derive analytical results for a weakly-anharmonic qutrit. Following this, numerical simulations with realistic anharmonicity values are used to corroborate these outcomes. We, again, simplify the Hamiltonian of Eq.~\eqref{eq:4-QutritRotFrameFull} by imposing a set of RWA conditions that read as
\begin{subequations}\label{eq:4-RWAConditionsThreeLevels}
\begin{align}
    &\omega_{\text{eg}}-\omega_{\text{r}} \ll \omega_{\text{eg}} +\omega_{\text{r}} \text{ and } \sqrt{2}g_1 \ll \min(\omega_{\text{eg}},\omega_{\text{r}}),\label{eq:4-RWAConditionsThreeLevels1} \\
    &\sqrt{2}g_1\ll 2\omega_{\text{d}}, \text{ and}\label{eq:4-RWAConditionsThreeLevels2}\\
    &\sqrt{2}\Omega_1\ll 4\omega_{\text{d}}.
    \label{eq:4-RWAConditionsThreeLevels3}
\end{align}
\end{subequations}
The conditions of Eq.~\eqref{eq:4-RWAConditionsThreeLevels1} are the usual conditions needed to arrive at a qutrit JCM Hamiltonian. The condition in Eq.~\eqref{eq:4-RWAConditionsThreeLevels2} is necessary to eliminate the counter-rotating interaction terms, whereas the condition in Eq.~\eqref{eq:4-RWAConditionsThreeLevels3} is required to drop the counter-rotating driving terms. Under all these RWA conditions, the simplified Hamiltonian reads as
\begin{align}
    \widehat{H}_{\text{RWA}}^{\text{d}}=&\frac{\hbar \Omega_1}{2}\left(\hat{\sigma}_{1+}+\hat{\sigma}_{1-} +\sqrt{2}\hat{\sigma}_{2+}+\sqrt{2}\hat{\sigma}_{2-}\right) + \hbar \delta \hat{a}^\dagger \hat{a}\nonumber\\
    &+\hbar g_1\left( \hat{\sigma}_{1+} +\sqrt{2} \hat{\sigma}_{2+}\right)\hat{a}\nonumber\\ &+\hbar g_1\left( \hat{\sigma}_{1-} +\sqrt{2} \hat{\sigma}_{2-}\right)\hat{a}^\dagger.
\end{align}
We now diagonalize the qutrit part of the free Hamiltonian, ${\hbar \Omega_1}(\hat{\sigma}_{1+}+\hat{\sigma}_{1-} +\sqrt{2}\hat{\sigma}_{2+}+\sqrt{2}\hat{\sigma}_{2-})/2$, to find the eigenvalues and eigenstates:
\begin{subequations}\label{eq:4-HarmonicThreeLevelSys}
\begin{align}
        \lambda_0=0,\,\ket{v_0}=\frac{1}{\sqrt{3}}\left(-\sqrt{2}\ket{\text{g}} + \ket{\text{f}}\right), \text{ and}
    \end{align}
    \begin{align}
        \lambda_{\pm}=\pm\frac{\hbar \Omega_1 \sqrt{3}}{2},\,\ket{v_\pm}=\frac{1}{\sqrt{3}}\left(\frac{1}{\sqrt{2}}\ket{\text{g}} \pm \sqrt{\frac{3}{2}}\ket{\text{e}} + \ket{\text{f}}\right).
    \end{align}
\end{subequations}
The zero eigenvalue state, $\ket{v_0}$, is commonly referred to as a \textit{dark state} in quantum optics \cite{Shore_CoherentAE}. In this system, this qutrit state does not get populated by the classical drive nor does it exchange photons with the resonator. This can also be seen as a mathematical feature of the zero eigenvalue, which makes it not evolve in time.
\begin{figure*}[t]
\includegraphics{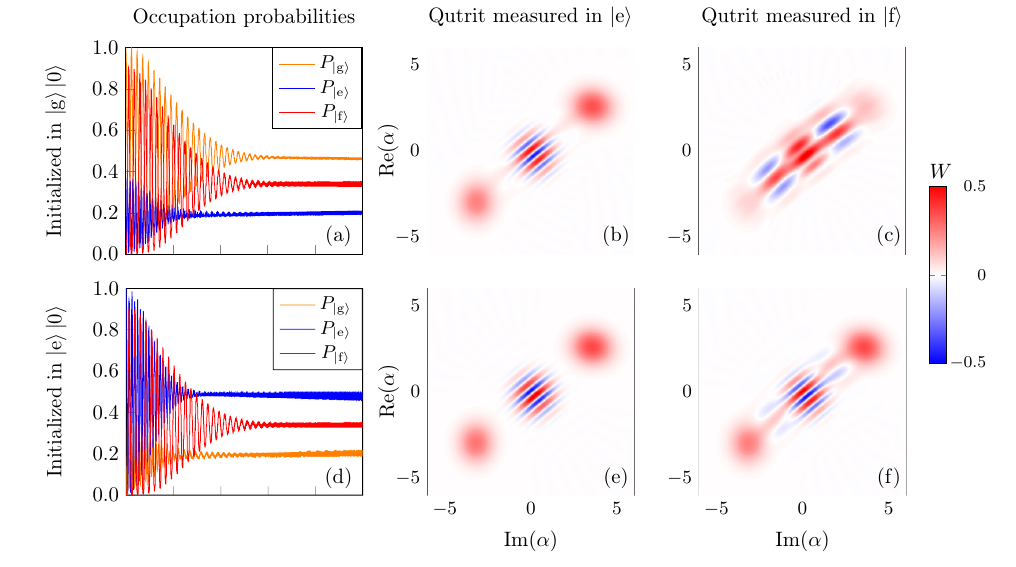}
\caption{\label{fig:Qutrit_g0_e0} Dynamics of cat states in a driven qutrit-resonator for different initial states. The parameters used for the simulations are: $\Omega_1=2\pi\times \SI{1}{\giga\hertz}$. $\Omega_2=\sqrt{2}\Omega_1$. $g_1=2\pi \times \SI{20}{\mega\hertz}$. $g_2=\sqrt{2}g_1$. $\Delta_1=\delta=0$. $\omega_{\text{q}}=2\pi\times \SI{5}{\giga\hertz}$. $\xi=-2\pi\times \SI{100}{\mega\hertz}$. $\gamma_1=\kappa=\SI{500}{\kilo\hertz}$. $\gamma_2=2 \gamma_1$. $\gamma_\phi= \SI{1}{\mega\hertz}$. The Wigner functions are obtained by projectively measuring the qutrit after a time-evolution period of $g_1 t /2\pi = 0.61$. (a)-(c) The dynamics of the system when the initial state is $\ket{\text{g}}\ket{0}$. (d)-(f) The dynamics of the system when the initial state is $\ket{\text{e}}\ket{0}$. When the qutrit is measured in $\ket{\text{g}}$, the resonator is found in a state very similar to that when the qutrit is measured in $\ket{\text{f}}$ (see Eqs.~\eqref{eq:QutritState_g0} and~\eqref{eq:QutritState_e0}).  } 
\end{figure*}
\twocolumngrid

Now that we found the dressed qutrit states, we follow the same procedure as in the qubit case by proceeding to the interaction picture. We define $\widehat{H}_0^{\text{d}}=\hbar \Omega_1\left(\hat{\sigma}_{1+}+\hat{\sigma}_{1-} \sqrt{2}\hat{\sigma}_{2+}+\sqrt{2}\hat{\sigma}_{2-}\right)/2 + \hbar \delta \hat{a}^\dagger \hat{a}$ and $\widehat{H}_{\text{I}}^{\text{d}}=\hbar g_1( \hat{\sigma}_{1+} +\sqrt{2} \hat{\sigma}_{2+})\hat{a}+\hbar g_1( \hat{\sigma}_{1-} +\sqrt{2} \hat{\sigma}_{2-})\hat{a}^\dagger$. Then, the interaction picture Hamiltonian reads as
\begin{align}\label{eq:4-FullIntPicHam}
    \widehat{H}^{(\text{I})}=&\hbar g_1 \Bigg[ \frac{\sqrt{3}}{2}(\dyad{v_+}-\dyad{v_-}) \nonumber \\ &-\frac{\sqrt{3}}{6}\dyad{v_+}{v_-}e^{i\Omega_1 \sqrt{3} t} + \frac{\sqrt{3}}{6}\dyad{v_-}{v_+}e^{-i\Omega_1 \sqrt{3} t} \nonumber\\  &-\sqrt{4} \dyad{v_+}{v_0}e^{i\Omega_1 \sqrt{3}t/2} + \sqrt{4}\dyad{v_-}{v_0}e^{-i\Omega_1 \sqrt{3}t/2} \nonumber\\ &+ \frac{1}{\sqrt{3}}\dyad{v_0}{v_+}e^{-i\Omega_1 \sqrt{3}t/2}\nonumber\\ & -\frac{1}{\sqrt{3}} \dyad{v_0}{v_-}e^{i\Omega_1 \sqrt{3}t/2} \Bigg]\hat{a}e^{-i\delta t} + \text{H.c.}.
\end{align}

Similar to the qubit case, the Hamiltonian of Eq.~\eqref{eq:4-FullIntPicHam} comprises two distinct interactions: a time-independent (diagonal) interaction and a time-dependent-drive-modulated (off-diagonal) interaction. We now impose the strong driving condition, $g_1,|\delta| \ll \Omega_1$, which when combined with the driving RWA condition of Eq.~\eqref{eq:4-RWAConditionsThreeLevels3} becomes \footnote{{Note that the qutrit bound is tighter than the qubit bound in Eq.~\eqref{eq:FullStrongDrivCond}.}}
\begin{align}
    g_1,\,|\delta| \ll \Omega_1 \ll 4 \omega_{\text{d}}/\sqrt{2}.
\end{align}
This allows us to neglect the drive-modulated terms and obtain the effective Hamiltonian
\begin{align}
    \widehat{H}^{(\text{I})}_{\text{eff}}= \hbar g_1\frac{ \sqrt{3}}{2}\left( \dyad{v_+}-\dyad{v_-} \right)\left( \hat{a}^\dagger e^{i\delta t} + \hat{a}e^{-i\delta t}\right).
\end{align}
This Hamiltonian generalizes that of Eq.~\eqref{eq:CatHam} for our qutrit case. Interestingly, it generates displacement in the resonator conditioned on two of the three qutrit dressed basis states, $\ket{v_+}$ and $\ket{v_-}$.

We now analyze the dynamics stemming from the qubit recipe for generating a cat state, where the initial state is  $\ket{\psi_i}=\ket{\text{g}}\ket{0}$. We rewrite the qutrit state using the dressed basis $\{\ket{v_0},\ket{v_+},\ket{v_-}\}$, and we find that $\ket{\psi_i}=\ket{\text{g}}\ket{0}=(c_{\text{g}0}\ket{v_0}+c_{\text{g}+}\ket{v_+}+c_{\text{g}-}\ket{v_-})\ket{0},$ where $c_{\text{g}0}=-\sqrt{2/3}$ and $c_{\text{g}\pm}={1}/{\sqrt{6}}.$ Then, the interaction picture time-evolved state is
\begin{align}\label{eq:QutritState_g0}
    \ket{\psi(t)}^{\text{(I)}}&=-\sqrt{\frac{2}{3}}\ket{v_0}\ket{0}+\frac{1}{\sqrt{6}}\ket{v_+}\ket{\alpha}+\frac{1}{\sqrt{6}}\ket{v_-}\ket{-\alpha}\nonumber\\
    &= \frac{1}{6}\ket{\text{g}}\left(4\ket{0}+\ket{\alpha} + \ket{-\alpha}\right)\nonumber\\
    &\,\,\,\,\,\,\,+ \frac{1}{\sqrt{12}}\ket{\text{e}}\left(\ket{\alpha} - \ket{-\alpha}\right)\nonumber\\
    &\,\,\,\,\,\,\,+ \frac{1}{\sqrt{18}}\ket{\text{f}}\left(2\ket{0}+\ket{\alpha} + \ket{-\alpha}\right),
\end{align}
where $\alpha=\sqrt{3}g_1(e^{i\delta t}-1)/2\delta$; when $\delta\rightarrow 0$, then $\alpha=-i\sqrt{3}gt/2$. Since any term involving the dark state remains unchanged in the time-evolved state. We find the resonator vacuum state $\ket{0}$ coupled to $\ket{\text{g}}$ and $\ket{\text{f}}$ (does not couple to $\ket{\text{e}}$, since $\braket{\text{e}}{v_0}=0$). Measuring the qutrit in $\ket{\text{e}}$ yields an odd cat state in the resonator, exactly as in the qubit case. {When measuring the qutrit in either $\ket{\text{g}}$ or $\ket{\text{f}}$, another non-classical state within the resonator, which is a superposition of the vacuum state and an even cat state, obtained; $\bra{\text{g}}\ket{\psi(t)}^{\text{(I)}}\propto 4\ket{0}+\ket{\alpha} + \ket{-\alpha}$ and $\bra{\text{f}}\ket{\psi(t)}^{\text{(I)}}\propto 2\ket{0}+\ket{\alpha} + \ket{-\alpha}$.}

{
This class of states is intriguing in its own right, exhibiting varying Wigner-negative regions and interference patterns (see Fig.~\ref{fig:Qutrit_g0_e0}(c)). While the goal of this section is to investigate cat states using qutrits, we briefly comment on the potential uses of this class of states. When $|\alpha|\gg 1$, the overlap between the states $\{\ket{0},\ket{\alpha},\ket{-\alpha}\}$ becomes very small, and the states are quasiorthogonal. As a result, we can encode a \textit{logical qutrit} in the resonator, with each of the three states corresponding to a logical qutrit state. This encoding generalizes the qubit cat code to the case of a qutrit. Additionally, these states can be employed to generate approximate Gottesman-Kitaev-Preskill (GKP) states \cite{GKP} by squeezing the state orthogonally to the axis of displacement \cite{SqueezedComb}. The resulting state exhibits a Gaussian profile, as required in finite-energy approximations of GKP states \cite{BosonicQIPRev,QECGKP}.}

The presence of the dark state prevents us from deterministically encoding a qubit state in the resonator using cat states. This is because measuring the qutrit in $\ket{\text{g}}$ or $\ket{\text{f}}$ does not leave a cat state in the resonator. To maximize the probability of finding a cat state in the resonator, we seek an alternative recipe. Initializing the qutrit in $\ket{\text{e}}$ removes $\ket{v_0}$ and its stationary vacuum contribution due to its decoupling from the dark state subspace. Therefore, we propose a new recipe for a generating a cat state tailored to the qutrit. Let the system start in an initial state $\ket{\psi_i}=\ket{\text{e}}\ket{0}=(\ket{v_+}-\ket{v_-})\ket{0}/\sqrt{2}$. Then, the interaction picture time-evolved state for this initial state is
\begin{align}\label{eq:QutritState_e0}
    \ket{\psi(t)}^{\text{(I)}}&=\frac{1}{\sqrt{2}}\left(\ket{v_+}\ket{\alpha}-\ket{v_-}\ket{-\alpha}\right)\nonumber\\
    &= \left(\frac{1}{\sqrt{12}}\ket{\text{g}} + \frac{1}{\sqrt{6}}\ket{\text{f}}\right)\left(\ket{\alpha}-\ket{-\alpha}\right)\nonumber\\ &\,\,\,\,\,\,\,\,+ \frac{1}{2}\ket{\text{e}}\left(\ket{\alpha}+\ket{-\alpha}\right).
\end{align}

We note that a projective measurement on $\ket{\text{g}},\,\ket{\text{e}}$ and $\ket{\text{f}}$ leaves a cat state in the resonator. While the specified initial state yields a cat state, the parity remains conditional on the qutrit state -- exactly as in the qubit case. To encode a qubit state, $c_{\text{g}}\ket{\text{g}}+c_{\text{e}}\ket{\text{e}}$ , in a cat state as described in Sec.~\ref{sec:QubitSysHamCond}, prepare the system in an initial state $\ket{\psi_{\text{i}}}=(c_{\text{g}}\ket{v_+}+c_{\text{e}}\ket{v_-})\ket{0}$. After time-evolving for the desired period and measuring in the bare basis $\{\ket{\text{g}},\ket{\text{e}},\ket{\text{f}}\}$, the resonator is left in a state $\propto c_{\text{g}}\ket{\alpha}\pm c_{\text{e}}\ket{-\alpha} $. This serves as a generalized procedure for encoding a qubit state in a resonator using a driven qutrit-resonator system.

The aim of the qutrit extension is to model a weakly anharmonic system, e.g., the transmon. Up until now, our work has been based on the assumption of the qutrit's perfect harmonicity, which encompasses multiple assumptions. We previously justified the perfect harmonicity of the coupling strengths. As for the detuning between the transition frequencies, $|\xi|$, the typical values are on the order of $100-300\text{ MHz}$ for a weakly-anharmonic system such as a transmon. Next, we show show that the perfectly harmonic qutrit model serves as a very good approximation for weakly anharmonic qutrits.

Figure~\ref{fig:Qutrit_g0_e0} displays the results of numerical simulations of the complete system Hamiltonian of Eq.~(\ref{eq:4-QutritSysHam}), without any approximations and using nonzero anharmonicity. These simulations are performed in presence of both qutrit and resonator decoherence by means of a Lindblad master equation, as explained in App.~\ref{app:Decoherence}. Figure~\ref{fig:Qutrit_g0_e0}(a) shows the occupation probabilities of the qutrit states when the system is initialized in $\ket{\text{g}}\ket{0}$. The probability of finding a cat state, for this initial state, depends on measuring the qutrit in $\ket{\text{e}}$ -- which is low. The Wigner functions of the resonator state after a projective measurement on different qutrit states are shown in Fig.~\ref{fig:Qutrit_g0_e0}. The Wigner functions of the states shown match the analytical predictions. The anharmonicity used for the simulations is $\xi=-100$ MHz \footnote{We performed simulations using anharmonicity values between $50$ MHz and $400$ MHz. From $50$ MHz to $\sim200$ MHz, the qualitative predictions of the perfectly harmonic qutrit hold true. See App.~\ref{app:ArbAnharm} for considerations regarding arbitrary anharmonicity.}. This anharmonicity is easily achievable by the transmon. The effect of nonzero anharmoncity can be seen in the pertubed interference regions in some of the aforementioned Wigner functions. Additionally, the lobes of the cat states get slightly deformed, but this is also due to the terms neglected in the strong driving RWA.

\subsection{Strongly-anharmonic limit}\label{sec:StronglyAnharm}

In this section, we discuss the limit of large anharmonicity $|\xi|$, while maintaining the perfectly harmonic coupling strengths. When $|\xi|$ is very large compared to $\Omega_1$ and $g_1$, the second transition ($\ket{\text{e}}\leftrightarrow\ket{\text{f}}$) is very far detuned from the first transition ($\ket{\text{g}}\leftrightarrow\ket{\text{e}}$). As a result, the driving barely affects the second transition and the resonator is either decoupled from or dispersively coupled to it depending on how far detuned it is. In either case, the effective dynamics are those of a driven qubit-resonator system with slight perturbations to the cat state from small third state leakage population.

\begin{figure}[ht]
\includegraphics[scale=1]{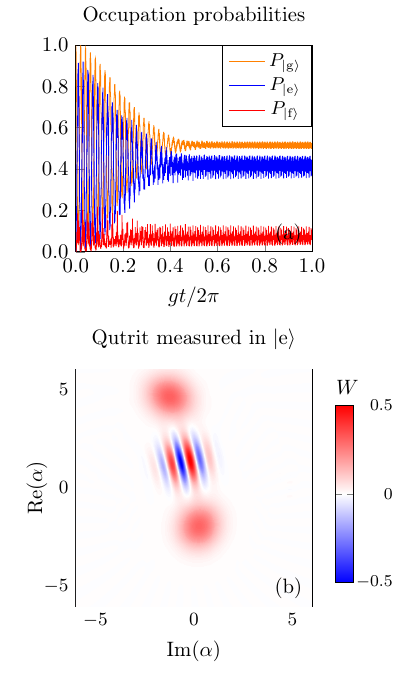}
        \caption{Strongly-anharmonic driven qutrit-resonator dynamics. The parameters used for the simulations are: $\Omega_1=2\pi\times \SI{1}{\giga\hertz}$. $\Omega_2=\sqrt{2}\Omega_1$. $g_1=2\pi \times \SI{20}{\mega\hertz}$. $g_2=\sqrt{2}g_1$. $\Delta_1=\delta=0$. $\omega_{\text{q}}=2\pi\times \SI{5}{\giga\hertz}$. $\xi=2\pi\times \SI{2}{\giga\hertz}$. $\gamma_1=\kappa=\SI{500}{\kilo\hertz}$. $\gamma_2=2 \gamma_1$. $\gamma_\phi= \SI{1}{\mega\hertz}$. The Wigner functions are obtained by projectively measuring the qutrit after a time-evolution period of $g_1 t /2\pi = 0.61$. The results resemble those of a driven qubit-resonator. The cat state is displaced off-center due to the contributions of the $\ket{\text{e}}\leftrightarrow\ket{\text{f}}$ transition.}\label{fig:LargeAnharmQutrit}
\end{figure}

In Fig.~\ref{fig:LargeAnharmQutrit}, we show the results of numerical simulations for such a regime. Figure~\ref{fig:LargeAnharmQutrit}(a) shows the occupation probabilities with the ${\ket{\text{f}}}$ population fluctuating around $0.1$ ($10\%$). Note that the sign of $\xi$ does not change the dymanics for a fixed set of selection rules. We can justify these arguments analytically by explicitly performing an RWA that eliminates terms oscillating with $e^{\pm i\widetilde{\chi}t}$ in the Hamiltonian of Eq.~\eqref{eq:4-QutritRotFrameFull} when $|\xi|\gg \Omega_1, g_1$.

We can discuss this regime of $|\xi|\gg\Omega_1,g_1$ in an experimental implementation context. For a charge qubit (operating at the degeneracy point), the anharmonicity between the two transitions fits this strongly-anharmonic regime. In this case, the system behaves as a driven qubit-resonator. This can be extrapolated to different circuit implementations such as flux-based circuits in the appropriate regimes. 

Up to this point, we have explored two extremes: the weakly- and strongly-anharmonic limits. Throughout our discussion, we have consistently upheld the harmonic assumption regarding the coupling strengths. However, for the intermediate regimes, we extend our analytical framework, as detailed in App.~\ref{app:ArbAnharm}, to address these scenarios while also relaxing the assumptions concerning the coupling strengths.

\section{Discussion}\label{sec:5}

The goal of this section serves two purposes: firstly, to contextualize our proposal in relation to other pertinent techniques commonly employed in circuit QED, and secondly, to explore the potential applicability of our method to qutrits with varying selection rules.

\subsection{Comparison to other methods}

We briefly listed methods used to generate cat states relevant to circuit QED in the introduction. The most relevant techniques to date are:~an engineered two-photon loss method \cite{DynProtecCat_2014,TwoPhotonLossCat}, a dispersive method known as `qcMAP' \cite{qcMAP,qcMAPexp} and a two-photon driven KNR \cite{TwoPhotonDrivCat,KerrCatExp}.

For the qcMAP, it depends on the ac-Stark shift, $\chi=g^2 / (\omega_{\text{q}}-\omega_{\text{r}})$, and thus is limited by $\chi$. The cat size for this method is $|\alpha_{\text{qcMAP}}(t)|^2\simeq 15/2(\chi t - \pi)$. {It is also worth noting that the dispersive conditional-displacement relies on the bare basis and, as a result, requires a further qubit rotation to place the coherent states in superposition.} Meanwhile, in the all-resonant case, our cat size is $|\alpha(t)|=g^2 t^2 /4$. This means that for similar parameters \footnote{The qcMAP operates in the dispersive regime, so either $\omega_{\text{r}}$ or $\omega_{\text{q}}$ has to be different from our resonant case.}, our proposal outperforms the qcMAP by generating cats of the same size in a much shorter time. The main premise of the qcMAP is the ability to deterministically encode a qubit state into a resonator. As shown in Sec.~\ref{sec:QubitSysHamCond}, our method also allows for the encoding of a qubit state in a resonator with a qubit-state-dependent parity, $c_{\text{g}}\ket{\text{g}}+c_{\text{e}}\ket{\text{e}}\mapsto c_{\text{g}}\ket{\alpha}\pm c_{\text{e}}\ket{-\alpha}$. For a single encoded qubit, this is not an issue since it can be tracked once the state is prepared. However, for multiple encoded qubits, this becomes an issue since they could have differing parities. A type of 'parity-fluid' protocol would need to be developed for this. Alternatively, a possible remedy would be to repeat the state preparation until the desired parity is obtained, enabled by the fast quadratic growth of the cat.

The two-photon driven KNR method depends on the Kerr nonlinearity of the system ($\hat{a}^{\dagger 2}\hat{a}^2$ with strength $\eta$) and the strength of the two-photon drive. Cat states are instantaneous degenerate eigenstates of the two-photon driven KNR, allowing for adiabatic preparation. Additionally, a counter-diabatic two-photon drive, as described in \cite{TransQuantumDriv_Berry}, can be used to accelerate the adiabatic state preparation while leveraging pulse optimization. In Ref.~\cite{TwoPhotonDrivCat}, no estimate of the cat size as a function of system parameters are provided. However, the authors showcase the ability to prepare cats with $|\alpha|^2 \approx 4$ photons on a timescale $\tau \simeq 1/\eta$ (accounting for reasonable single-photon loss $\kappa=\eta/250$). Typically, $\eta$ is at most on the order of 10 MHz, and for such $\eta$ values, the timescale is on the order of 100 nanoseconds. {Taking the resonator single-photon loss into consideration, we can approximate our cat size (in the all-resonant case) by $|\alpha(t)|^2\simeq g^2 t^2 e^{-\kappa t}/4$ (see App.~\ref{app:Decoherence} for the details on this estimate and further decoherence considerations).} This means that even for modest values of $g$, we can achieve the same size cat or even a larger one without the need for pulse optimization or counter-diabatic driving.

Lastly, the two-photon loss method partially overlaps with the two-photon driven KNR protocol. When a resonator is subjected to a two-photon drive and a two-photon dissipative process, odd Fock states converge to the odd cat states and even Fock states converge to the even cat state. The advantage is that the steady states of the system are the desired cat states. This is precisely the core of the two-photon loss scheme. We think that our protocol is complementary to this method. Explicitly, one can generate the cat state using our protocol and ensure its confinement via the two-photon loss scheme.

\subsection{Qutrits with different selection rules}

In our extension of the proposal to a driven qutrit-resonator system, we exclusively focused on the $\Xi$-type qutrit. However, it's worth noting that there are three other types of qutrits: $\Lambda$-type, $V$-type, and $\Delta$-type. For the $\Lambda$-type qutrit, the allowed transitions are $\ket{\text{g}}\leftrightarrow \ket{\text{f}}$ and $\ket{\text{e}}\leftrightarrow \ket{\text{f}}$. In the case of the $V$-type qutrit, the allowed transitions include $\ket{\text{g}}\leftrightarrow \ket{\text{f}}$ and $\ket{\text{g}}\leftrightarrow \ket{\text{e}}$. Lastly, for the $\Delta$-type qutrit, all transitions are allowed.

Our framework can be readily extended to accommodate $\Lambda$- and $V$-type qutrits. For a $\Lambda$-type qutrit, we replace the transition operators in Eq.~\eqref{eq:4-QutritSysHam} with $\hat{\sigma}_{1+}^{\Lambda}=\dyad{\text{f}}{\text{g}}$ and $\hat{\sigma}_{2+}^{\Lambda}=\dyad{\text{f}}{\text{e}}$. For a $V$-type qutrit, the replacement transition operators are $\hat{\sigma}_{1+}^{V}=\dyad{\text{e}}{\text{g}}$ and $\hat{\sigma}_{2+}^{V}=\dyad{\text{f}}{\text{e}}$. Additional work is necessary to account for a cyclic $\Delta$-type qutrit, and this will be addressed in a later work.

With these substitutions, if the coupling strengths also exhibit harmonic scaling, the findings presented in Secs~\ref{sec:WeaklyAnharm} and~\ref{sec:StronglyAnharm} can be directly applied. Otherwise, one can use the general framework outlined in App.~\ref{app:ArbAnharm}.

\section{Summary and Conclusions}\label{sec:6}

We introduced a cat-state generation method centered on a driven qubit-resonator system with linear coupling as its foundation. We described the requisite conditions and outlined the regime of validity, examining scenarios encompassing both resonant and detuned qubit drives. The method's operation is primarily situated in the strong driving regime, characterized by the dynamics of the driving Bloch-Siegert shift. This shift induces oscillations in the weighting coefficients of the cat lobes, a phenomenon that can be tracked. Subsequently, we leveraged this method to encode a qubit state within the resonator (with a qubit- and qutrit-state-dependent parity). We then generalized the method to a driven qutrit-resonator. We analyzed the regimes of weak and strong anharmonicities along with their implications for experimental implementation candidates, e.g., a transmon or charge qubit. Throughout, we showed the robustness of our protocol against qubit and resonator decoherence, using worse-than-average decoherence parameters in numerical simulations.

Next, we situated our method within the context of cat-state generation protocols commonly employed in circuit QED and demonstrated its adaptability to qutrits with varying selection rules, including $\Lambda$-type and $V$-type qutrits. 

Finally, we believe that a fast scheme generating cat states that grow quadratically in time using a resonant or detuned qubit drive, as proposed here, presents a valuable tool for all purposes of generating a cat state.

\begin{acknowledgments}
This research was undertaken thanks in part to funding from
the Canada First Research Excellence Fund (CFREF). We
acknowledge the support of the Natural Sciences and Engineering Research Council of Canada (NSERC), [Application
Number: RGPIN-2019-04022].
\end{acknowledgments}


\appendix

\section{Decoherence}\label{app:Decoherence}

\begin{figure*}[t]
\includegraphics[trim={1cm 0 0 0},scale=1.1]{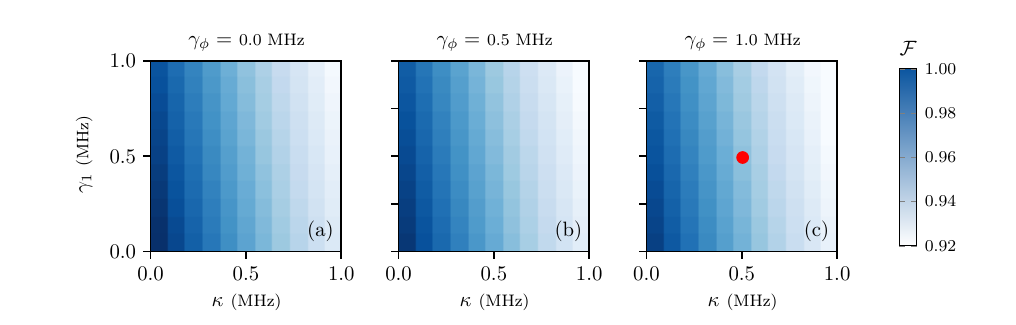}
\caption{\label{fig:Fidelity} {Fidelity of prepared state with varying decoherence rates. The Hamiltonian parameters and initial state used for the simulations are identical to those in Fig.~\ref{fig:QubitResDynamics} with the decoherence parameters varied. (a)-(c) The fidelity between a state prepared via time-evolution using Eq.\eqref{eq:QubitMasterEqn} and a reference state evolved with Eq.\eqref{eq:SysHam} is plotted for different qubit and resonator decoherence rates, with a time-evolution period of $gt/2\pi=1.0$. The resonator photon loss rate, $\kappa$, is the most detrimental parameter to the state fidelity. Meanwhile, the qubit relaxation rate, $\gamma_1$, diminishes the fidelity but the qubit dephasing, $\gamma_\phi$, is practically negligible as the three plots are nearly identical. The red dot in (c) represents the decoherence parameters used in Fig.~\ref{fig:QubitResDynamics} and Fig.~\ref{fig:QubitDetDynamics}.}} 
\end{figure*}
\twocolumngrid

The numerical simulations are performed using a Lindblad master equation at zero-temperature, assuming the qubit-resonator system interact with their environment baths separately. The master equation for the qubit-resonator system reads as \cite{Breuer_Petruccione_2010,Carmichael_1993}
\begin{align}\label{eq:QubitMasterEqn}
    \frac{d}{dt}\hat{\rho}=-\frac{i}{\hbar}[\widehat{H},\hat{\rho}] + \gamma_1 \mathcal{D}(\hat{\sigma}_{-})\hat{\rho} +  \frac{\gamma_{\phi}}{2} \mathcal{D}(\hat{\sigma}_z)\hat{\rho} + \kappa \mathcal{D}(\hat{a})\hat{\rho},
\end{align}
where $\hat{\rho}$ is the full system density matrix, $\mathcal{D}(\hat{O})\hat{\rho}=\hat{O}\hat{\rho}\hat{O}^\dagger - \{\hat{O}^\dagger \hat{O},\hat{\rho}\}/2$ is the dissipator for a given operator $\hat{O}$, $\gamma_1$ and $\gamma_\phi$ are the qubit energy relaxation and dephasing rate, and $\kappa$ is the resonator photon loss rate. While for the qutrit-resonator system, the master equation reads as
\begin{align}
    \frac{d}{dt}\hat{\rho}=&-\frac{i}{\hbar}[\widehat{H},\hat{\rho}] + \gamma_1 \mathcal{D}(\hat{\sigma}_{1-})\hat{\rho} + \gamma_2 \mathcal{D}(\hat{\sigma}_{2-})\hat{\rho}\nonumber\\ &+  \frac{\gamma_{\phi}}{2} \mathcal{D}(\hat{\sigma}_z+ 2 \dyad{\text{f}})\hat{\rho} + \kappa \mathcal{D}(\hat{a})\hat{\rho}.
\end{align}
Here, $\gamma_2$ is the energy relaxation rate of the second transition.
For the purpose of simulating a qutrit resembling a transmon, we assume $\gamma_2\simeq 2 \gamma_1$ and that the dephasing operator is $\hat{\sigma}_z + 2\dyad{\text{f}}.$ Explicitly, this is related to the fact that in a harmonic oscillator, the n$^\text{th}$ Fock state has a relaxation rate $\gamma_n=n\gamma_1$, $n$-times the relaxation rate of the first excited Fock state \cite{FockStateDecay}. Additionally, experiments testing the coherence times of higher levels of transmons (even beyond $\ket{\text{f}}$) have confirmed this Fock-state-like scaling of decay times \cite{TransmonHigherDecay}. These assumptions are quite reasonable for both transmon and charge qubits. All the numerical simulations of the master equation were performed using QuTiP \cite{Johansson_Nation_Nori_2012, Johansson_Nation_Nori_2013}.

{We can obtain an approximate analytic estimate for the resonator's photon number in the presence of single-photon loss since $\alpha \mapsto \alpha e^{-\kappa t/2}$ \cite{QuantumOpticsKlimov_Chumakov_2009}. Thus, in the case of an open system, specifically in the scenario of all-resonant conditions, $|\alpha|^2 = g^2 t^2 e^{-\kappa t}/4$ ($|\widetilde{\alpha}|^2 = g^2 \Omega^2 t^2 e^{-\kappa t}/4\varepsilon^2$ in the cross-resonant case where $\Delta\neq 0$ and $\delta=0$). We can now use this expression to find the maximum size a cat can reach by differentiating it with respect to time and equating the derivative to zero. The time at which $\alpha$ reaches a maximum is $t_{\text{max}}=2/\kappa$ (valid for both all-resonant and cross-resonant cases). This is a rough estimate based on the resonator single-photon loss channel.}

{For a full quantitative picture, we further analyze the system by varying the decoherence parameters to determine its effect on the fidelity of the prepared state, $\hat{\rho}_{\text{prep}}$. We take the fidelity to be defined as $$\mathcal{F}=\left(\Tr(\sqrt{\sqrt{\hat{\rho}_{\text{prep}}}\hat{\rho}_{\text{ideal}}\sqrt{\hat{\rho}_{\text{prep}}}})\right)^2,$$
where we use a reference ideal state $\hat{\rho}_{\text{ideal}}$ arrived at only using the system Hamiltonian in Eq.~\eqref{eq:SysHam} and $\hat{\rho}_{\text{prep}}$ is arrived at using the master equation, Eq.~\eqref{eq:QubitMasterEqn}. Figure~\ref{fig:Fidelity} illustrates the outcomes of numerical simulations in which the reference and prepared states underwent time evolution over a normalized period of $gt/2\pi=1$. In the same figure, it is evident that the resonator photon loss rate emerges as the most detrimental factor influencing the state fidelity. Although qubit relaxation also contributes to fidelity reduction, the impact of qubit dephasing is almost negligible within the considered timescale. Notably, the maximum decoherence rates employed in our simulations, set at $\SI{1}{\mega\hertz}$, are considerably more severe than those typically encountered in current circuit QED setups, let alone state-of-the-art devices. }

\section{Deformed cat states}

\begin{figure}[t]
\includegraphics[scale=1]{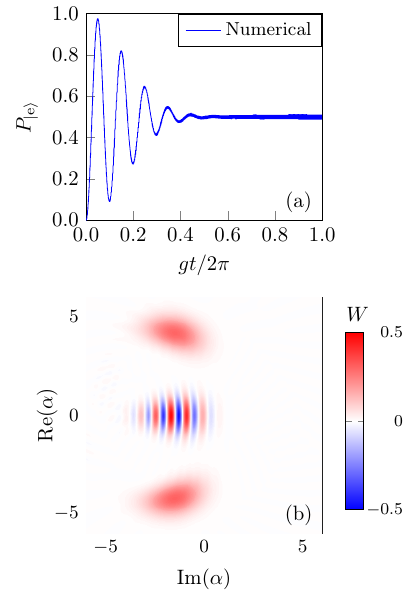}
        \caption{Deformed cat state dynamics. The parameters used for the simulations are: $\Omega=2\pi\times \SI{200}{\mega\hertz}$. $\Delta=\delta=0$. $\omega_{\text{q}}=2\pi\times \SI{5}{\giga\hertz}$. $g=2\pi \times \SI{20}{\mega\hertz}$. $\gamma_1=\kappa=\SI{500}{\kilo\hertz}$. $\gamma_\phi= \SI{1}{\mega\hertz}$. The Wigner function $W$ of the resonator state is obtained after measuring the qubit in $\ket{\text{e}}$ at $gt/2\pi= 1$. The cat state exhibits squeezing of the lobes and interference region and a displacement on both real and imaginary axes. }\label{fig:DeformedCat}
\end{figure}

The driving regimes considered in the main text have all be centered around strong driving. Although, a cat state can be achieved with a weaker drive, i.e., when $\Omega \gg |\delta|,g$ does not hold. An example of such state is shown in Fig.~\ref{fig:DeformedCat}. This cat state is manifestly \emph{deformed}. To understand the origin of the deformation, we group together all the terms dropped to obtain the Hamiltonian of Eq.~\eqref{eq:CatHam} in what we define the deformation Hamiltonian:

\begin{align}\label{eq:DeformedHam}
\widehat{H}^{(\text{I})}_{\text{def}}&=\widehat{H}_{\text{RWA}}^{\text{(I)}}-\widehat{{H}}_{\text{eff}}^{\text{(I)}}\nonumber \\ &=\frac{\hbar g}{2}\left(e^{i \Omega t} \dyad{+}{-} - e^{-i\Omega t}\dyad{-}{+}\right)\hat{a}e^{-i\delta t} + \text{H.c.} \nonumber\\
&= \frac{i\hbar g}{2}\left[\hat{\sigma}_y\cos(\Omega t)+\hat{\sigma}_z\sin(\Omega t)\right]\left(e^{+i\delta t}\hat{a}^\dagger-e^{-i\delta t}\hat{a}\right).
\end{align}

The dynamics described by the deformation Hamiltonian also exhibit a conditional displacement, similar to Eq.~\eqref{eq:CatHam}. However, in this case, the displacement is modulated by the classical drive and is conditioned on two different qubit operators, $\hat{\sigma}_z$ and $\hat{\sigma}_y$. Furthermore, the deformation Hamiltonian introduces a displacement in a different direction in phase space (specifically, on the real axis) compared to the displacement generated by Eq.~\eqref{eq:CatHam}. In the presence of a weak drive, the contributions from $\widehat{H}^{(\text{I})}_{\text{def}}$ become nonnegligible. When measuring in the bare basis of the qubit, $\{\ket{\text{g}},\ket{\text{e}}\}$ (as required to obtain the cat state given by Eq.~\eqref{eq:CatState}), we observe a cat state that exhibits properties resembling squeezing.

\section{Spurious resonator drive}\label{sec:Spurious}

When considering the implementation scheme of a driven qubit-resonator system, we must account for the cross-talk between the qubit drive and the resonator. This cross-talk results in a spurious driving of the resonator. When the qubit is coupled to the resonator and the drive capacitively, the system is composed of a three-port capacitive network, and the couplings are made explicit in the capacitance matrix
\begin{equation}
\label{eq:Cmatrix}
    \boldsymbol{C}=\begin{pNiceMatrix}[first-row,first-col]
    & \text{qubit} & \text{resonator} & \text{drive}       \\
\text{qubit} & C_{11}   & C_{12}   & C_{13}\\
\text{resonator} & C_{21}   & C_{22}   & C_{23}\\
\text{drive}  & C_{31}   & C_{32}    & C_{33}\\
\end{pNiceMatrix}.
\end{equation}
The capacitance matrix is taken to be real and symmetric, i.e., $C_{ij}=C_{ji}\in\mathbb{R}$. To obtain the capacitive part of the circuit QED Hamiltonian, we typically resort to \cite{cQEDRev}
\begin{align}
    \widehat{H}_{\text{cap}}=\frac{1}{2}\hat{\vec{Q}}^{T}\boldsymbol{C}^{-1}\hat{\vec{Q}},
\end{align}
where $\hat{\vec{Q}}$ is the vector of charge operators and $\boldsymbol{C}^{-1}$ is the inverse of the capacitance matrix in Eq.~\eqref{eq:Cmatrix}. It is worth explicitly expressing the inverse capacitance matrix to elaborate on the presence of a spurious coupling even when there is no direct capacitive coupling. We point out that $(C^{-1})_{13}= (C_{12}C_{23}-C_{13}C_{22})/\det\boldsymbol{C}$, which determines the coupling between the drive and resonator in the Hamiltonian, can have a nonzero value even when $C_{13}=0$.

The spurious resonator drive modifies the Hamiltonian of Eq.~\eqref{eq:SysHam} to
 \begin{align}
     \widehat{H}'= &\frac{\hbar \omega_{\text{q}}}{2}\hat{\sigma}_z + \hbar \omega_{\text{r}} \hat{a}^\dagger \hat{a} + \hbar g (\hat{\sigma}_+ + \hat{\sigma}_-)(\hat{a}^\dagger + \hat{a}) \nonumber\\ &+ \hbar\Omega\cos(\omega_{\text{d}} t)(\hat{\sigma}_+ + \hat{\sigma}_-) \nonumber\\ &+ \hbar{\Omega'}\cos(\omega_{\text{d}} t+{\phi'})(\hat{a}^\dagger + \hat{a}),
     \label{eq:SysHamSpurious}
 \end{align}
where the spurious drive is characterized by a strength, $\Omega'$, and a phase, $\phi'$. The presence of this drive displaces the resonator state by an additional coherent state. To address this, we add a cancellation drive to the resonator that is physically far from the qubit such that its effects are minimal on the qubit. The Hamiltonian of the system now reads as 
\begin{align}
     \widehat{H}'= &\frac{\hbar \omega_{\text{q}}}{2}\hat{\sigma}_z + \hbar \omega_{\text{r}} \hat{a}^\dagger \hat{a} + \hbar g (\hat{\sigma}_+ + \hat{\sigma}_-)(\hat{a}^\dagger + \hat{a}) \nonumber\\ &+ \hbar\Omega\cos(\omega_{\text{d}} t)(\hat{\sigma}_+ + \hat{\sigma}_-) \nonumber\\ &+ \hbar({\Omega'}\cos(\omega_{\text{d}} t+{\phi'})+{\Omega_{\text{c}}}\cos(\omega_{\text{d}} t+{\phi}_{\text{c}}))(\hat{a}^\dagger + \hat{a}),
     \label{eq:SysHamSpurious2}
 \end{align}
 where $\Omega_{\text{c}}$ and $\phi_{\text{c}}$ are the strength and phase of the cancellation drive. Then, we can experimentally find $\Omega'$ and $\phi'$ by 
identifying the coherent state displacing the resonator state. Finally, setting $\Omega_{\text{c}}=\Omega'$ and $\phi_{\text{c}}=\phi'+\pi$ and using the identity $\cos(x+\pi)=-\cos(x)$, we achieve total cancellation of the spurious drive.

\section{Extension to an arbitrarily-anharmonic qutrit}\label{app:ArbAnharm}

In this section, we extend our analytical framework to address arbitrarily anharmonic qutrits by carefully selecting an appropriate rotating frame. Additionally, we relax assumptions on coupling strengths and assume arbitrary coupling strengths. However, we maintain the bare minimum assumption, $g_1/g_2=\Omega_1/\Omega_2$. This is because the mechanism in which the qutrit physically couples to the driving field is the same as it does to the resonator.

We start by transforming driven qutrit-resonator Hamiltonian of Eq.~\eqref{eq:4-QutritSysHam} into a different rotating frame. This rotating frame is defined by the unitary transformation $\hat{U}=\exp[-it\left(\omega_{\text{d}}\left(\dyad{\text{e}}-\dyad{\text{g}} - \dyad{\text{f}}\right)  + 2\omega_{\text{d}}\hat{a}^\dagger \hat{a}\right)/2] $. In this frame, the system Hamiltonian reads as

\begin{align}\label{eq:4-RotFrame2}
    \widehat{H}^{\text{d}}=& \frac{\hbar \Delta_1}{2}\left(\dyad{\text{e}} - \dyad{\text{g}} \right)+ \frac{\hbar \Sigma}{2}\dyad{\text{f}} + \hbar\delta \hat{a}^\dagger \hat{a}\nonumber\\
    &+  \frac{\hbar \Omega_1}{2}\left(\hat{\sigma}_{1+}+\hat{\sigma}_{1-}+e^{i2\omega_{\text{d}}t}\hat{\sigma}_{1+} +e^{-i2\omega_{\text{d}}t}\hat{\sigma}_{1-}\right)\nonumber\\
    &+  \frac{\hbar \Omega_2}{2}\left(\hat{\sigma}_{2+}+\hat{\sigma}_{2-}+e^{i2\omega_{\text{d}}t}\hat{\sigma}_{2+} +e^{-i2\omega_{\text{d}}t}\hat{\sigma}_{2-}\right)\nonumber\\
    &+\hbar\bigg[ g_1\left(e^{i\omega_{\text{d}}t}\hat{\sigma}_{1+} + e^{-i\omega_{\text{d}}t}\hat{\sigma}_{1-}\right)
    \nonumber\\ &\,\,\,\,\,\,\,\,\,\,\,\,\,\,\,+g_2\left(e^{i\omega_{\text{d}}t}\hat{\sigma}_{2+} + e^{-i\omega_{\text{d}}t}\hat{\sigma}_{2-}\right)\bigg]\nonumber\\
    &\times\left(e^{i\omega_{\text{d}}t}\hat{a}^\dagger + e^{-i\omega_{\text{d}}t}\hat{a} \right),
\end{align}
where $\Sigma=\widetilde{\omega}_{\text{f}}+ \omega_{\text{d}}.$ The intuition behind this particular frame is that it sets both transitions on equal footing. The off-diagonal terms have the same form in their time-dependence for both transitions. This Hamiltonian can be simplified by a set of RWA conditions that read
\begin{subequations}\label{eq:4-RWAConditions_ArbAnharm}
\begin{align}
    &\omega_{\text{eg}}-\omega_{\text{r}} \ll \omega_{\text{eg}} +\omega_{\text{r}} \text{ and } g_1 \ll \min(\omega_{\text{eg}},\omega_{\text{r}}),\label{eq:4-RWAConditions_ArbAnharm1} \\
    &g_1\ll 2\omega_{\text{d}},\label{eq:4-RWAConditions_ArbAnharm2}\\
    &\Omega_1\ll 4\omega_{\text{d}},
    \label{eq:4-RWAConditions_ArbAnharm3}\\
    &\omega_{\text{fe}}-\omega_{\text{r}} \ll \omega_{\text{fe}} +\omega_{\text{r}} \text{ and } g_2 \ll \min(\omega_{\text{fe}},\omega_{\text{r}}),\label{eq:4-RWAConditions_ArbAnharm4} \\
    &g_2\ll 2\omega_{\text{d}},\text{ and}\label{eq:4-RWAConditions_ArbAnharm5}\\
    &\Omega_2\ll 4\omega_{\text{d}}.
    \label{eq:4-RWAConditions_ArbAnharm6}
\end{align}
\end{subequations}
These conditions generalize those of Eq.~\eqref{eq:4-RWAConditionsThreeLevels} for arbitrary coupling strengths. Then, assuming all the conditions stated above allows us to perform an RWA and obtain the Hamiltonian
\begin{align}
    \widehat{H}^{\text{d}}_{\text{RWA}}=& \frac{\hbar \Delta_1}{2}\left(\dyad{\text{e}} - \dyad{\text{g}} \right)+ \frac{\hbar \Sigma}{2}\dyad{\text{f}} + \hbar\delta \hat{a}^\dagger \hat{a}\nonumber\\
    &+  \frac{\hbar \Omega_1}{2}\left(\hat{\sigma}_{1+}+\hat{\sigma}_{1-}\right)
    +  \frac{\hbar \Omega_2}{2}\left(\hat{\sigma}_{2+}+\hat{\sigma}_{2-}\right)\nonumber\\
    &+ \hbar\left( g_1 \hat{\sigma}_{1+}+g_2 \hat{\sigma}_{2+}\right)\hat{a} + \hbar\left( g_1 \hat{\sigma}_{1-}+g_2 \hat{\sigma}_{2-}\right)\hat{a}^\dagger.
\end{align}
For simplicity, we assume $\Delta_1=0.$ We now diagonalize the qutrit part of the free Hamiltonian, $\hbar \Sigma \dyad{\text{f}}/2 + \hbar \Omega_1 (\hat{\sigma}_{1+} + \hat{\sigma}_{1-})/2 +\hbar \Omega_2 (\hat{\sigma}_{2+} + \hat{\sigma}_{2-})/2$, to find the eigenvalues and eigenstates \cite{Shore_CoherentAE}:
\begin{align}
    &\lambda_1=-\frac{1}{3}a+\frac{2}{3}p\cos\left(\frac{\theta}{3}\right),\\ &\lambda_2=-\frac{1}{3}a -\frac{2}{3}p\cos\left(\frac{\theta}{3}+\frac{\pi}{3}\right),\\ &\lambda_3=-\frac{1}{3}a -\frac{2}{3}p\cos\left(\frac{\theta}{3}-\frac{\pi}{3}\right),
\end{align}
and
\begin{align}
    \ket{v_k}=&\frac{1}{\mathcal{N}_k} \Bigg[ \left(\Omega_1 \left(\lambda_k - \frac{\Sigma}{2}\right)\right) \ket{\text{g}}\nonumber\\&\,\,\,\,\,\,\,\,\,\,\,\,+ \left(2\lambda_k\left(\lambda_k-\frac{\Sigma}{2}\right)\right)\ket{\text{e}} + \Omega_2\ket{\text{f}}\Bigg],
\end{align}
where
\begin{subequations}
\begin{align}
    &a=-\frac{\Sigma}{2},\\
    &b=-\frac{1}{4}\left(\Omega_1^2 + \Omega_2^2\right),\\
    &c=\frac{1}{8}\Sigma \Omega_2^2,\\
    &p=\sqrt{a^2 -3b},\\
    &\cos\theta=-\frac{27c + 2a^3 -9ab}{2p^3},
\end{align}
and
\begin{align}
\mathcal{N}_k=\sqrt{\lambda_k^2|\Omega_2|^2 + (4\lambda_k^2 + |\Omega_1|^2)\left(\lambda_k-\frac{\Sigma}{2}\right)^2}
\end{align}
\end{subequations}
for $k=1,2,3$. Next, we proceed as done in previous sections by transforming to the interaction picture. We define $\widehat{H}_0= {\hbar \Sigma}\dyad{\text{f}}/2 + \hbar\delta \hat{a}^\dagger \hat{a}
    +  {\hbar \Omega_1}(\hat{\sigma}_{1+}+\hat{\sigma}_{1-})/2
    +  {\hbar \Omega_2}(\hat{\sigma}_{2+}+\hat{\sigma}_{2-})/2$ and $ \widehat{H}_{\text{I}}=\hbar( g_1 \hat{\sigma}_{1+}+g_2 \hat{\sigma}_{2+})\hat{a} + \hbar( g_1 \hat{\sigma}_{1-}+g_2 \hat{\sigma}_{2-})\hat{a}^\dagger$. Then, the interaction picture Hamiltonian reads as
    \begin{widetext}
\begin{align}
    \widehat{H}^{(\text{I})}=&\hbar \sum_{k=1}^3 \Bigg[\frac{2  \lambda_k \left(\lambda_k - \frac{\Sigma}{2}\right)}{\mathcal{N}_k^2}\left(g_1\Omega_1 \left(\lambda_k - \frac{\Sigma}{2}\right) + g_2\Omega_2\right)\dyad{v_k}\nonumber\\
    &\,\,\,\,\,\,\,\,\,\,\,\,\,\,\,+ \sum_{l\neq k}\frac{\left(\lambda_l-\frac{\Sigma}{2}\right)}{\mathcal{N}_k\mathcal{N}_l}\left(2g_1\Omega_1 \lambda_k\left(\lambda_k-\frac{\Sigma}{2}\right)+g_2\Omega_2\right)\dyad{v_k}{v_l}e^{i\Delta \lambda_{kl}t}\Bigg] \left( \hat{a}^\dagger e^{+i\delta t} + \hat{a} e^{-i\delta t} \right),
\end{align}
\end{widetext}
where $\Delta \lambda_{kl}=\lambda_k -\lambda_l$. In this Hamiltonian, there time-independent (diagonal) terms and time-dependent (off-diagonal) terms. The separation of timescales cannot be assumed \textit{a priori}. To achieve a similar Hamiltonian as in the previous sections, we must find the \textit{strong driving-anharmoncity} \footnote{The parameters $\Delta\lambda_{kl}$ are functions of the driving strengths $\Omega_1$ and $\Omega_2$, as well as $\Sigma$, which can be directly reformulated in terms of $\xi$.} regime where \begin{align}
     g_1,g_2,|\delta|\ll |\Delta\lambda_{kl}|.
\end{align}
We note that this has to be satisfied with the RWA conditions of Eq.~\eqref{eq:4-RWAConditions_ArbAnharm}.
If this regime is achieved, we can neglect the off-diagonal terms modulated by $\Delta\lambda_{kl}$ and obtain the effective Hamiltonian
\begin{align}
    \widehat{H}^{(\text{I})}_{\text{eff}}=&\hbar \sum_{k=1}^3 \widetilde{g}_k\dyad{v_k} \left( \hat{a}^\dagger e^{+i\delta t} + \hat{a} e^{-i\delta t} \right),
\end{align}
where $\widetilde{g}_k=2  \lambda_k (\lambda_k - \Sigma/2)(g_1\Omega_1 (\lambda_k - {\Sigma}/{2}) + g_2\Omega_2)/{\mathcal{N}_k^2}.$ This Hamiltonian generates resonator displacements conditioned the dressed qutrit basis $\{\ket{v_1},\ket{v_2},\ket{v_3}\}.$ Let the initial state be $\ket{\psi_{\text{i}}}=\sum_{k=1}^3 c_k \ket{v_k}\ket{0}$ with $\sum_{k=1}^3 |c_k|^2=1$. Then, the time-evolved state under the effective Hamiltonian yields
\begin{align}
    \ket{\psi(t)}^{(\text{I})}= \sum_{k=1}^3 c_k\ket{v_k}\ket{\alpha_k},
\end{align}
where $\alpha_k=-\widetilde{g}_k(e^{i\delta t}-1)/2\delta$; when $\delta\rightarrow 0$, then $\alpha_k=-i\widetilde{g}_kt/2$. Similar to the previous sections, we can create interesting nonclassical states composed of a superposition of coherent states by rewriting the states in the bare basis and measuring the qutrit. This leaves the resonator in a nonclassical state dependent on which qutrit state was measured. 

The framework presented in this section has been quite general and abstracted away from a particular circuit implementation. The purpose of this section is to present a general framework for an arbitrarily anharmonic driven qutrit-resonator system that does not conform to either of the two extremes introduced in the previous two sections. Additionally, the approach presented here is useful for tailoring parameters to a particular device. One can use the derived analytical eigenvalues and eigenstates along with all the given constraints to numerically optimize for the generation of a cat state or a particular (collinear \footnote{All the displacements generated are on the same axis.}) superposition of coherent states of interest for a set of particular device parameters. 

\bibliography{paper}

\end{document}